\documentclass[universe,article,accept,moreauthors,pdftex]{Definitions/mdpi} 

\graphicspath{{./Definitions/}}
\usepackage{relsize}
\usepackage{blindtext}
\usepackage{enumitem}
\usepackage{graphicx}
\usepackage{bm}
\usepackage{cancel}
\usepackage{epsfig}
\usepackage{dcolumn}
\usepackage{amsmath}
\usepackage{hhline}
\usepackage{adjustbox}
\usepackage{enumerate}
\usepackage[utf8]{inputenc}
\usepackage{caption}
\usepackage{longtable}

\def\aaps{{Astron.\@  \@Astroph.\ }}
\def\apss{{Astroph.\@ \& \@Space \@Science\ }}
\def\apjl{{Astroph. J. Lett.}}
\def\apj{{Astroph.\@ J.\ }}

\def\nat{{Nature\ }}

\def \beq  {\begin{equation}}
\def \eeq  {\end{equation}}
\def \ber  {\begin{eqnarray}}
\def \eer  {\end{eqnarray}}

\firstpage{1} 
\pubvolume{1}
\issuenum{1}
\articlenumber{0}
\pubyear{2021}
\copyrightyear{2021}
\externaleditor{Academic Editor: Lorenzo Iorio} 

\datereceived{2 September 2021} 
\dateaccepted{22 September 2021} 
\datepublished{} 
\hreflink{https://doi.org/10.3390/\linebreak universe7100366} 

\Title{HINTS FOR A GRAVITATIONAL TRANSITION IN {TULLY--FISHER DATA}} 

\TitleCitation{Hints for a Gravitational Transition in the  Tully-Fisher Data}

\Author{George Alestas \orcidA{}, Ioannis Antoniou \orcidB{} and Leandros Perivolaropoulos *\orcidC{}}

\AuthorNames{George Alestas, Ioannis Antoniou and Leandros Perivolaropoulos}

\address[1]{%
Department of Physics, University of Ioannina, 45110 Ioannina, Greece;  g.alestas@uoi.gr (G.A.); \linebreak  i.antoniou@uoi.gr (I.A.)}

\AuthorCitation{Alestas, G.; Antoniou, I.; Perivolaropoulos, L.}

\corres{\hangafter=1 \hangindent=1.05em \hspace{-0.82em}Correspondence: leandros@uoi.gr}

\newcommand{\newc}{\newcommand}

\newcommand{\be}{\begin{equation}}
\newcommand{\ee}{\end{equation}}
\newcommand{\ba}{\begin{eqnarray}}
\newcommand{\ea}{\end{eqnarray}}
\newcommand{\bea}{\begin{eqnarray*}}
\newcommand{\eea}{\end{eqnarray*}}
\newc{\D}{\partial}
\newc{\ie}{{i.e.,} }
\newc{\eg}{{e.g.,} }
\newc{\etc}{{etc.} }
\newc{\etal}{{et al.}}
\newc{\lcdm}{$\Lambda$CDM }
\newc{\lcdmnospace}{$\Lambda$CDM}
\newc{\wcdm}{$w$CDM }
\newc{\plcdm}{Planck18/$\Lambda$CDM }
\newc{\plcdmnospace}{Planck18/$\Lambda$CDM}
\newc{\omom}{$\Omega_{0m}$ }
\newc{\omomnospace}{$\Omega_{0m}$}

\newc{\ra}{\Rightarrow}
\newc{\baodv}{$\frac{D_V}{r_s}$ }
\newc{\baodvnospace}{$\frac{D_V}{r_s}$}
\newc{\baoda}{$\frac{D_A}{r_s}$ } 
\newc{\baodanospace}{$\frac{D_A}{r_s}$}
\newc{\baodh}{$\frac{D_H}{r_s}$ }
\newc{\baodhnospace}{$\frac{D_H}{r_s}$}

\abstract{We use an up-to-date compilation of Tully--Fisher data to search for transitions in the evolution of the Tully--Fisher relation. Using an up-to-date data compilation, we find hints at $\approx 3\sigma$ level for a transition at critical distances $D_c \simeq 9$  Mpc and $D_c \simeq 17$  Mpc. We split the full sample in two subsamples, according to the measured galaxy distance with respect to splitting distance $D_c$, and identify the likelihood of the best-fit slope and intercept of one sample with respect to the best-fit corresponding values of the other sample. For $D_c \simeq 9$  Mpc and $D_c \simeq 17$  Mpc, we find a tension between the two subsamples at a level of $\Delta \chi^2 > 17\;  (3.5\sigma)$. Using Monte Carlo simulations, we demonstrate that this result is robust with respect to random statistical and systematic variations of the galactic distances and is unlikely in the context of a homogeneous dataset constructed using the Tully--Fisher relation. If the tension is interpreted as being due to a gravitational strength transition, it would imply a shift in the effective gravitational constant to lower values for distances larger than $D_c$ by $\frac{\Delta G}{G}\simeq -0.1$. Such a shift is of the anticipated  sign and magnitude but at a somewhat lower distance (redshift) than the gravitational transition recently proposed to address the Hubble and growth tensions ($\frac{\Delta G}{G}\simeq -0.1$ at the transition redshift  of $z_t\lesssim 0.01$ ($D_c\lesssim 40$  Mpc)).}


\begin{document}
\section{Introduction}

The Tully–Fisher relation (TFR) \cite{Tully:1977fu} has been proposed as an empirical relation that connects the intrinsic optical luminosity of spiral galaxies with their observed maximum velocity $v_{rot}$ in the rotation curve as follows:
\be
L=A v_{rot}^s
\label{tfr}
\ee
where  $s\simeq 4$ is the slope in a logarithmic plot of \eqref{tfr}, and $A$ is a constant ($log(A)$ is the zero point or intercept). The~constants $s$ and $A$ appear to depend very weakly on  galaxy properties, including the mass to light ratio, the~observed surface brightness, the~galactic profiles, $HI$ gas content, size, \etc \cite{refId0}. They clearly also depend, however, on the fundamental properties of gravitational interactions as demonstrated~below. 

The baryonic Tully--Fisher relation (BTFR) is similar to Equation~\eqref{tfr} but connects the galaxy's total baryonic mass (the sum of mass in stars and $HI$ gas) $M_B$ with the rotation velocity as follows:
\be
M_B=A_B v_{rot}^s
\label{btfr}
\ee
where $A_B\simeq 50 M_\odot$ km$^{-4}$ s$^4$ \cite{McGaugh_2005}. This allows to include gas-rich  dwarf galaxies that appear in groups and have stellar masses below $10^9 M_{\odot}$.

A simple heuristic analytical derivation for the BTFR can be obtained as follows~\cite{1979ApJ...229....1A}. Consider a star in a circular orbit of radius $R$ around a galactic mass $M$ rotating with velocity $v$. Then, the following holds:
\be
v^2=G_{\rm eff}M/R\implies v^4=(G_{\rm eff}M/R)^2\sim M \; S\; G_{\rm eff}^2
\label{v2m}
\ee
where $G_{\rm eff}$ is the effective Newton's constant involved in gravitational interactions and $S$ the surface density $S\equiv M/R^2$, which is expected to be constant~\cite{1970ApJ...160..811F}. From \mbox{Equations \eqref{btfr} and \eqref{v2m}},  the following is anticipated:
\be
A_B\sim G_{\rm eff}^{-2} S^{-1}
\label{abg}
\ee
Therefore, the BTFR can, in principle, probe both galaxy formation dynamics (through, \eg~$S$) and possible fundamental constant dynamics (through $G_{\rm eff}$). An~interesting feature of the BTFR is that despite the above heuristic derivation, it appears to be robust, even in cases when the galaxy sample includes low $S$ and/or varying $S$ galaxies~\cite{Zwaan:1995uu, McGaugh:1998tq}. In~fact, no other parameter appears to be significant in the~BTFR.

The BTFR has been shown to have lower scatter~\cite{10.1111/j.1365-2966.2012.21469.x,10.1093/mnras/stw2461,refId0} than the classic stellar TFR and also to be applicable for galaxies with stellar masses lower than $10^9M_\odot$. It is also more robust than the classic TFR~\cite{1999Ap&SS.269..119F, 2000ApJ...533L..99M, 2001ApJ...563..694V, Zaritsky:2014dca} since the parameters $A_B$ (intercept) and $s$ (slope) are very weakly dependent on galactic properties, such as size and surface brightness~\cite{refId0}. 

The low scatter of the BTFR and its robustness make it useful as a distance indicator for the measurement of the Hubble constant $H_0$. A~calibration of the BTFR using Cepheid and TRGB distances leads to a value of $H_0=75\pm 3.8$ km s$^{-1}$ Mpc$^{-1}$ \cite{Schombert:2020pxm}. 

\textls[-15]{This value of $H_0$ is consistent with local measurements of $H_0$, using SnIa calibrated with Cepheids ($H_0=73.2\pm 1.3$ km s$^{-1}$ Mpc$^{-1}$) \cite{Riess:2020fzl}, but is in tension with the value of $H_0$ obtained using the early time sound horizon standard ruler calibrated using the CMB anisotropy spectrum in the context of the standard \lcdm model (\mbox{$H_0=67.36\pm 0.54$ km  s$^{-1}$ Mpc$^{-1}$})}~\cite{Aghanim:2018eyx}. The~tension between the CMB and Cepheid calibrators is at a level larger than $4\sigma$ and constitutes a major problem for modern cosmologies (for a recent review and approaches see Refs.~\cite{Perivolaropoulos:2021jda,DiValentino:2021izs, Kazantzidis:2019dvk, Alestas:2020mvb, 10.1093/mnras/stab1070}). 

The Hubble tension may also be viewed as an inconsistency between the value of the standardized SnIa absolute magnitude $M$ calibrated using Cepheids in the redshift range $0<z<0.01$ (distance ladder calibration) and the corresponding $M$ value calibrated using the recombination sound horizon (inverse distance ladder calibration) for $0.01<z<z_{rec}$. Thus, a~recently proposed class of approaches to the resolution of the Hubble tension involves a transition~\cite{Marra:2021fvf, Alestas:2020zol} of the standardized intrinsic SnIa luminosity $L$ and absolute magnitude $M$ at a redshift $z_t\lesssim 0.01$ from $M=(-19.24\pm 0.037)\; mag$ for $z<z_t$ (as implied by Cepheid calibration) to $M=(-19.4\pm 0.027)\; mag$ for $z>z_t$ (as implied by CMB calibration of the sound horizon at decoupling) \cite{Camarena:2019rmj}. Such a transition may occur due to a transition in the strength of the gravitational interactions $G_{\rm eff}$, which modifies the SnIa intrinsic luminosity $L$ by changing the value of the Chandrasekhar mass. The~simplest assumption leads to $L\sim M_{Ch}\sim G_{\rm eff}^{-3/2}$  \cite{Amendola:1999vu,Gaztanaga:2001fh}, even though corrections may be required to the above simplistic approach~\cite{Wright:2017rsu}.

The weak evolution and scatter of the BTFR can be used as a probe of galaxy formation models as well as a probe of possible transitions of fundamental properties of gravitational dynamics since the zero point constant $A_B$ is inversely proportional to the square of the gravitational constant $G$. Previous studies investigating the evolution of the best-fit zero point $log\; A_B$ and slope $s$ of the BTFR have found a mildly high $z$ evolution of the zero point from $z\simeq 0.9$ to $z\simeq 2.3$ \cite{2017ApJ...842..121U}, which was attributed to the galactic evolution inducing a lower gas fraction at low redshifts after comparing with the corresponding evolution of the stellar TFR (STFR), which ignores the contribution of gas in the galactic~masses.

Ref.~\cite{2017ApJ...842..121U} and other similar studies assumed a fixed strength of fundamental gravitational interactions and made no attempt to search for sharp features in the evolution of the zero point. In~addition, they focused on the comparison of high redshift with low redshift effects without searching for possible transitions within the low $z$ spiral galaxy data. Such transitions, if present, would be washed out and hidden from these studies, due to averaging effects. In~the present analysis, we search for transition effects in the BTFR at $z\lesssim 0.01$ (distances $D\lesssim 40\rm \,Mpc$), which may be due to either astrophysical mechanisms or to a rapid transition in the strength of the gravitational interactions $G_{\rm eff}$, due to fundamental~physics.  

In many modified gravity theories, including scalar tensor theories, the strength of gravitational interactions $G_{\rm eff}$ measured in Cavendish-type experiments measuring force $F$ between masses ($F=G_{\rm eff}\frac{m_1\; m_2}{r^2}$), is distinct from the Planck mass corresponding to $G_N$ that determines the cosmological background expansion rate ($H^2 =\frac{8\pi G_N}{3}\rho_{tot}$). 

For example, in~scalar tensor theories involving a scalar field $\phi$ and a non-minimal coupling $F(\phi)$ of the scalar field to the Ricci scalar in the Lagrangian, the~gravitational interaction strength is as follows~\cite{EspositoFarese:2000ij}:
\be
G_{\rm eff}=\frac{1}{F}\frac{2F+4F,_{\phi}^2}{2F+3F,_{\phi}^2}
\label{geffst}
\ee
while the Planck mass related $G_N$ is as follows:
\be
G_{N}=\frac{1}{F}
\label{gnst}
\ee

Most current astrophysical and cosmological constraints on Newton's constant constrain the time derivative of $G_{\rm eff}$ at specific times,~assume a smooth power--law evolution of $G_{\rm eff}$, or constrain changes of the Planck mass--related $G_N$ instead of $G_{\rm eff}$ (CMB and nucleosynthesis constraints~\cite{Alvey:2019ctk}).  Therefore, these studies are less sensitive in the detection of rapid transitions of $G_{\rm eff}$ at low $z$. 

The current constraints on the evolution of $G_{\rm eff}$ and $G_N$ are summarized in Table~\ref{table1}, where we review the experimental constraints from local and cosmological time scales on the time variation of the gravitational constant. The methods are based on very diverse physics, and the resulting upper bounds differ by several orders of magnitude. Most constraints are obtained from systems in which gravity is non-negligible, such as the motion of the bodies of the solar system, and the astrophysical and cosmological systems. They are mainly related in the comparison of a gravitational time scale, \eg~period of orbits, with~a non-gravitational time scale. One can distinguish between two types of constraints, from~observations on cosmological scales and on local (inner galactic or astrophysical) scales. The~strongest constraints to date come from lunar ranging~experiments.

\begin{specialtable}
\caption{Solar system, astrophysical and cosmological constraints on the evolution of the gravitational constant. Methods with star (*) constrain $G_N$, while the rest constrain $G_{\rm eff}$. The~latest and strongest constraints are shown for each~method.}

\label{table1}
\begin{adjustbox}{width=\linewidth,left}
\begin{tabular}{ccccc}  \toprule
  \textbf{Method}  & \boldmath{$\Big|\frac{\Delta G_{\rm eff}}{G_{\rm eff}}\Big|_{max}$} & \boldmath{$\Big|\frac{\dot G_{\rm eff}}{G_{\rm eff}}\Big|_{max}$ \textbf{($yr^{-1}$)}} & \textbf{Time Scale (Yr) }&  \textbf{References} \\
  \midrule 
Lunar ranging &  & $1.47\times 10^{-13}$ & 24 &\cite{Hofmann:2018myc} \\  
Solar system &  & $4.6\times 10^{-14}$ & 50 &\cite{Pitjeva:2021hnc,Pitjeva:2013xxa}  \\  
Pulsar timing &  &$3.1\times 10^{-12}$ &1.5 &\cite{Deller:2008jx} \\ 
Strong Lensing &  &$10^{-2}$ &0.6 &\cite{Giani:2020fpz} \\ 
Orbits of binary pulsar & &$1.0\times 10^{-12}$ &22 &\cite{Zhu:2018etc}  \\ 
Ephemeris of Mercury & &$4\times 10^{-14}$ &7 &\cite{PMID:29348613}\\ 
Exoplanetary motion & &$ 10^{-6}$  &4 &\cite{Masuda:2016ggi} \\ 
Hubble diagram SnIa & 0.1 & $1\times 10^{-11}$&$\sim$$10^8$ &\cite{Gazta_aga_2009}\\ 
Pulsating white-dwarfs & &$1.8\times 10^{-10}$  & 0&\cite{Corsico:2013ida}\\ 
Viking lander ranging & &$4\times 10^{-12}$ & $6$ & \cite{PhysRevLett.51.1609} \\ 
Helioseismology & & $1.6\times 10^{-12}$&$4\times 10^9$ &\cite{Guenther_1998} \\ 
Gravitational waves&$8$ &$5\times10^{-8}$ & $1.3\times 10^{8}$ &\cite{Vijaykumar:2020nzc}  \\ 
Paleontology & $0.1$&$2\times 10^{-11}$ &$4\times 10^9$ &\cite{Uzan:2002vq}  \\ 
Globular clusters & &$35\times 10^{-12}$ & $\sim$$10^{10}$&\cite{DeglInnocenti:1995hbi}  \\ 
Binary pulsar masses &  &$4.8\times 10^{-12}$&$\sim$$10^{10}$ &\cite{Thorsett:1996fr} \\ 
Gravitochemical heating & &$4\times 10^{-12}$ &$\sim$$10^8$ &\cite{Jofre:2006ug} \\ 
Strong lensing & &$3\times 10^{-1}$ &$\sim$$10^{10}$ &\cite{Giani:2020fpz} \\ 
Big Bang Nucleosynthesis *  &$0.05$ &$4.5\times 10^{-12}$ &$1.4\times 10^{10}$ &\cite{Alvey:2019ctk}\\ 
Anisotropies in CMB * &$0.095$ &$1.75\times 10^{-12}$ &$1.4\times 10^{10}$ &\cite{Wu:2009zb}\\ \bottomrule
\end{tabular}
\end{adjustbox}

\end{specialtable}

In the first column of Table~\ref{table1}, we list the used method. The~second column contains the upper bound $\big|\frac{\Delta G}{G}\big|_{max}$ of the fractional change of $G$ during the corresponding timescale. Most of these bounds assume a smooth evolution of $G$.  In~the third column, we present the upper bound on the normalized time derivative  $\big|\frac{\dot G}{G}\big|_{max}$. The~fourth column is an approximate time scale over which each experiment is averaging each variation, and the fifth column refers to the corresponding study where the bound appears. Entries with a star ($*$) indicate constraints on $G_N$, while the rest of the constraints refer to the gravitational interaction constant $G_{\rm eff}$. 

In the present analysis, we search for a transition of the BTFR best-fit parameter values (intercept and slope) between data subsamples at low and high distances. We consider sample dividing distances $D_c\in [2,60]\rm\,Mpc$, using a robust BTFR dataset~\cite{2001ApJ...563..694V,Walter_2008,2019MNRAS.484.3267L,2016ApJ...816L..14L}, 
which consists of 118 carefully selected BTFR datapoints, providing distance, rotation velocity baryonic mass ($D,V_f,M_B$) as well as other observables with their $1\sigma$ errorbars.   We  focus on the gravitational strength Newton constant $G_{\rm eff}$ and address the following~questions:
\begin{itemize}
\item Are there hints for a transition in the evolution of the BTFR? 
\item
What constraints can be imposed on a possible $G_{\rm eff}$ transition, using BTFR data?
\item
Are these constraints consistent with the level of $G_{\rm eff}$ required to address the Hubble~tension?
\end{itemize}

The structure of this paper is the following: In the next section, we  describe the datasets involved in our analysis and present the method used to identify transitions in the evolution of the BTFR at low $z$. We also show the results of our analysis. In~Section~\ref{sec3}, we summarize, present our conclusions and discuss possible implications and extensions of our~analysis.

\section{Search for Transitions in the Evolution of the~BTFR}

The logarithmic form of the BTFR (Equation (\ref{btfr})) is as follows:
\be
y=log M_B = s \, log v_{rot} + log A_B \equiv s\; x + b
\label{logbtfr}
\ee
and a similar form for the TFR. Due to Equation~(\ref{abg}), the~intercept $b\equiv log A_B$ depends on both the galaxy formation mechanisms through the surface density $S$ and on the strength of gravitational interactions through $G_{\rm eff}$. 

A controversial issue in the literature is the type of possible evolution of the slope and intercept of the TFR and the BTFR. Most studies have searched for possible evolution in high redshifts  (redshift range $z\in [0,3]$) with controversial results. For~example, several studies found no statistically significant evolution of the intercept of the TFR up to redshifts of $z\sim 1.7$ \cite{Conselice:2005kt, Kassin:2007zz, Miller:2011nt, Contini, DiTeodoro, 2017MNRAS.466..892M, Pelliccia}, while other studies found a negative evolution of the intercept up to redshift $z\simeq 3$ \cite{Puech2008, Puech2010, Cresci2009Apr, Gnerucci2011, Swinbank:2012nz, Price_2016, 10.1093/mnras/stw936, Straatman_2017}. Similar controversial results in high $z$ appeared for the BTFR, where~\cite{Puech2010} found no significant evolution of the intercept since $z\simeq 0.6$, while~\cite{Price_2016} found a positive evolution of the intercept between low-z galaxies and a $z \simeq 2$ sample. In~addition, cosmological simulations of disc galaxy formation based on cosmological N-body/hydrodynamical simulations have indicated no evolution of the TFR based on stellar masses in the range $z\in [0.1]$ \cite{Portinari:2006wg}, indicating also that any observed evolution of the TFR is an artifact of the luminosity~evolution.

These studies have focused mainly on comparing high-$z$ with low-$z$ samples, making no attempt to scan low redshift samples for abrupt transitions of the intercept and slope. Such transitions would be hard to explain in the context of known galaxy formation mechanisms but are well motivated in the context of fundamental gravitational constant transitions, which may be used to address the Hubble tension~\cite{Alestas:2020zol, Marra:2021fvf}. Thus, in~this section, we attempt to fill this gap in the~literature. 

\textls[-18]{We consider the BTFR dataset shown in Appendix \ref{secA} based on the data from~\cite{2001ApJ...563..694V,Walter_2008,2019MNRAS.484.3267L,2016ApJ...816L..14L}} of the flat rotation velocity of galaxies vs. the baryonic mass (stars plus gas) consisting of 118 datapoints, shown in Table~\ref{table3}. The~sample is restricted to those objects for which both quantities are measured to better than $20\%$ accuracy and includes galaxies in the approximate distance range $D\in [1, 130]\rm\,Mpc$. This is a robust low $z$ dataset ($z<0.1$) with low scatter showing no evolution of velocity residuals as a function of the central surface density of the stellar~disks. 

Our analysis is distinct from previous studies in two~aspects:
\begin{itemize}
\item 
We use an exclusively low $z$ sample to search for BTFR evolution.
\item
We focus on a particular type of evolution: \textit{sharp transitions of the intercept and slope}.
\end{itemize}

\textls[-15]{In this context, we use the dataset shown in Table~\ref{table3} of Appendix \ref{secA}~\cite{2001ApJ...563..694V,Walter_2008,2019MNRAS.484.3267L,2016ApJ...816L..14L}}, consisting of the distance $D$, the~logarithm of the baryonic mass $log M_B$ and the logarithm of the asymptotically flat rotation velocity $log v_{rot}$ of 118 galaxies along with $1\sigma$ errors. We fix a critical distance $D_c$ and split this sample in two subsamples $\Sigma_1$ (galaxies with $D<D_c$) and $\Sigma_2$ (galaxies with $D>D_c$). For~each subsample, we use the maximum likelihood method~\cite{10.5555/1403886} and perform a linear fit to the data setting $y_i=log (M_B)_i$, \mbox{$x_i=log (v_{rot})_i$}, while the parameters to fit are the slope $s$ and the intercept $b$ of Equation~(\ref{logbtfr}). Thus, for each sample $j$ ($j=0,1,2$ with $j=0$ corresponding to the full sample and $j=1,2$ corresponding to the two subsamples $\Sigma_1$ and $\Sigma_2$), we minimize the following:
\be
\chi_{j}^2(s,b)=\sum_{i=1}^{N_j}\frac{\left[y_i-(s_j\; x_i +b_j)\right]^2}{s_j^2 \sigma_{xi}^2 +\sigma_{yi}^2 + \sigma_s^2}
\label{chi2def}
\ee
with respect to the slope $s_j$ and intercept $b_j$. We fix the scatter to $\sigma_s=0.077$, obtained by demanding that $\frac{\chi_{0,min}^2}{N_0}=1$, where $\chi_{0,min}^2$ is the minimized value of $\chi^2$ for the full sample and $N_0$ is the number of datapoints of the full sample. We thus find the best fit values of the parameters $s_j$ and $b_j$, ($j=0,1,2$) and also construct the $1\sigma-3\sigma$ likelihood contours in the $s-b$ parameter space for each sample (full, $\Sigma_1$ and $\Sigma_2$) for a given value of $D_c$. We then evaluate the $\Delta \chi_{kl}^2(D_c)$ of the best fit of each subsample $k$, best fit with respect to the likelihood contours of the other subsample $l$. Using these values, we also evaluate the $\sigma$-distances ($d_{\sigma, kl}(D_c)$ and $d_{\sigma, lk}(D_c)$) and conservatively define the minimum of these $\sigma$-distances  as follows:
\be
d_\sigma (D_c)\equiv Min\left[d_{\sigma, 12}(D_c), d_{\sigma, 21}(D_c)\right]
\label{dcdef}
\ee
For example, for the $\sigma$-distance of the best fit of $\Sigma_1$ with respect to the likelihood contours of $\Sigma_2$, we have the following:
\be
\Delta \chi_{12}^2 (D_c)\equiv \chi_2^2(s_1,b_1)(D_c)-\chi_{2,min}^2(s_2,b_2)(D_c)
\label{dchi}
\ee
and $d_{\sigma, 12}$ is obtained as a solution of the following equation~\cite{10.5555/1403886}:
\be
\Delta \chi_{12}^2 =2\; Q^{-1}\left[\frac{M}{2},1-{\rm Erf}(\frac{d_{\sigma, 12}}{\sqrt{2}})\right]
\label{dsigdef}
\ee
where $Q^{-1}$ is the inverse regularized incomplete Gamma function, $M$ is the number of parameters to fit ($M=2$ in our case) and Erf is the error~function. 
\begin{figure}[H]
\includegraphics[width = 0.73\textwidth]{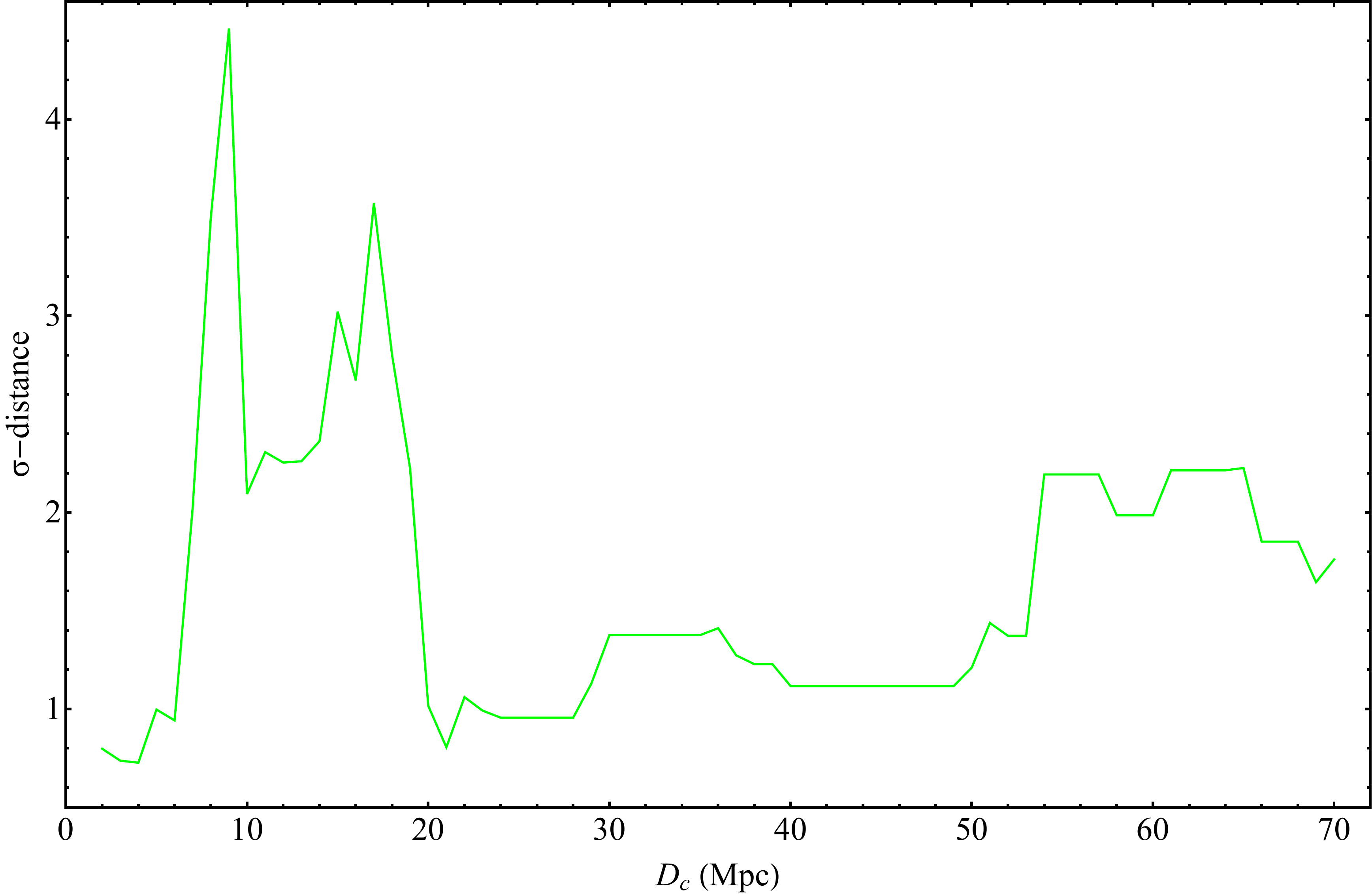}
\caption{The $\sigma$-distance between the various $\Sigma_1$ and $\Sigma_2$ datasets as a function of the split distances $D_c$. There are 2 clear peaks at $D_c = 9\;$  Mpc and $D_c = 17\;$  Mpc and a transition seems to have been completed at $D_c\simeq 20\rm \, Mpc$. The~anticipated plot would be a $\sigma$-distance that consistently varies in the range up to about $2\sigma$ for all values of $D_c$. The~observed peaks indicate either the presence of systematics or the presence of interesting~physics. }
\label{fig1}
\end{figure}
Figure~\ref{fig1} shows the $\sigma$ distance $d_\sigma(D_c)$ in the parameter space $(b,s)$ as a function of the split sample distance $D_c$. There are two peaks indicating larger than $3\sigma$ difference between the two subsamples at $D_c = 9\;$  Mpc and $D_c = 17\;$  Mpc. In~addition, a transition of the $\sigma$ distance $d_\sigma(D_c)$ at $D_c\simeq 20$  Mpc is apparent.
This Monte Carlo simulation is used to construct Figure~\ref{fig2_new} (right panel---green line---range), where we show the mean and standard deviation range of the $\sigma$-distances obtained by the above-described 100 Monte Carlo samples. Clearly, the random variation in the galactic distances cannot change the qualitative features (high double peak at low $D_c$) of Figure~\ref{fig1} corresponding to the real sample. The~$\sigma$-distances obtained from such a typical Monte Carlo sample is shown in Figure~\ref{fig2_new} (left panel green line).

\begin{figure}[H]
\hspace{-0.5cm}
\includegraphics[width = 0.78\textwidth]{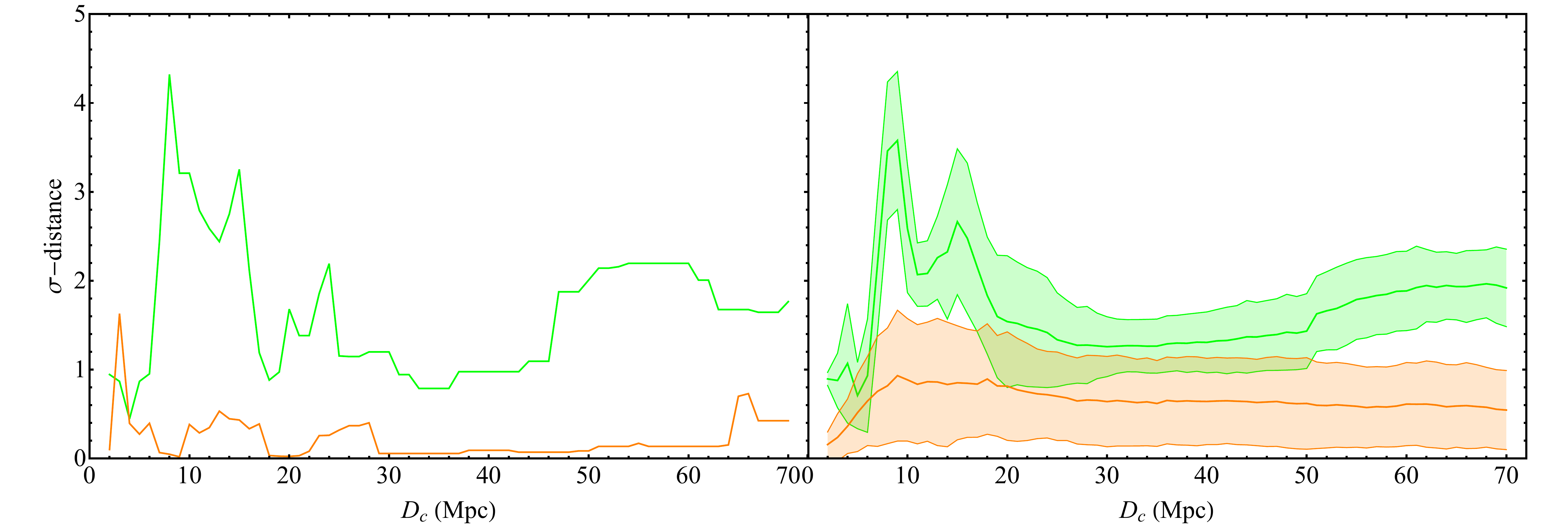}
\caption{{{\textbf{Left panel}}}: The $\sigma$-distances as a function of the split distances $D_c$ for a sample dataset with random distance values, normally distributed inside their individual $1-\sigma$ range (green line), versus the $\sigma$-distances as a function of the split distances $D_c$ for a homogeneous Monte Carlo sample constructed using the best-fit BTFR (orange line). {\bf{Right panel}}: The $68\%$ range of the $\sigma$-distances versus the split distances $D_c$ produced by a Monte Carlo simulation of 100 sample datasets obtained by randomly varying galaxy distance values with a Gaussian probability distribution (green band). Superimposed is the $68\%$ range of the $\sigma$-distances versus the split distances $D_c$ obtained from 100 homogeneous Monte Carlo samples constructed using the best-fit BTFR (orange band). Evidently, the~characteristic two-peak form of the plot remains practically unchanged, even after the random variation in the distances (green band), whereas no significant tension is present in the case of the homogeneous Monte Carlo samples for any value of $D_c$ (orange band).}
\label{fig2_new}
\end{figure}

\begin{specialtable}[H]

\caption{The best-fit values of the intercept and slope parameters corresponding to the likelihood contours of Figure~\ref{fig2} alongside with their $1\sigma$ errors. The~minimum $\Delta \chi^2$ between the best fits of the two samples is also shown. The~corresponding $\sigma$-tension in parenthesis is obtained in the context of two free parameters from Equation~(\ref{dsigdef}).  Notice that, even though the parameter values appear to be consistent, the~value of $\Delta \chi^2$ between the subsamples reveals the tension at $D_c=9\rm\, Mpc$ and $D_c=17$~Mpc. }
\label{table2}
\setlength{\cellWidtha}{\columnwidth/4-2\tabcolsep+0.0in}
\setlength{\cellWidthb}{\columnwidth/4-2\tabcolsep+0.0in}
\setlength{\cellWidthc}{\columnwidth/4-2\tabcolsep+0.0in}
\setlength{\cellWidthd}{\columnwidth/4-2\tabcolsep+0.0in}
\scalebox{1}[1]{\begin{tabularx}{\columnwidth}{>{\PreserveBackslash\centering}m{\cellWidtha}>{\PreserveBackslash\centering}m{\cellWidthb}>{\PreserveBackslash\centering}m{\cellWidthc}>{\PreserveBackslash\centering}m{\cellWidthd}}
\toprule

\boldmath$D_c$ \textbf{(Mpc)}  & \textbf{Intercept} & \textbf{Slope} & \boldmath{$\Delta \chi_{min}^2$} \\
\midrule 
- & $2.287 \pm 0.18$ & $3.7 \pm 0.08$ & - \\ \midrule
<9 & $2.461 \pm 0.407$ & $3.586 \pm 0.216$ & $23.7 \;  (4.5\sigma)$ \\ 
>9 & $2.854 \pm 0.379$ & $3.46 \pm 0.204$ & $23.7 \;  (4.5\sigma)$ \\ 
<17 & $2.467 \pm 0.38$ & $3.592 \pm 0.17$ & $17.0 \;  (3.7\sigma)$ \\ 
>17 & $2.677 \pm 0.368$ & $3.548 \pm 0.166$ & $17.0 \;  (3.7\sigma)$ \\ 
<40 & $2.327 \pm 0.987$ & $3.681 \pm 0.419$ & $2.9 \;  (1.2\sigma)$  \\ 
>40 & $3.318 \pm 0.816$ & $3.283 \pm 0.349$ & $2.9 \;  (1.2\sigma)$ \\ \bottomrule
\end{tabularx}}
\end{specialtable}

The typical qualitative feature of $d_\sigma(D_c)$ corresponding to the real sample disappears if we homogenize the sample by randomizing both the velocities and the galactic masses, using the measured values of the velocities and the estimated values of the galactic masses in the context of the best-fit BTFR. In~order to construct such a homogenized BTFR sample from the real sample, we use the following~steps:
\begin{itemize}
    \item 
    We assign to each galaxy a randomly chosen distance obtained from a Gaussian distribution with mean equal to the measured distance and standard deviation equal to the $1\sigma$ error of the measured distance.
    \item
    We assign to each galaxy a randomly chosen $logv_{rot} $ obtained from a Gaussian distribution with mean equal to the measured $logv_{rot} $ and standard deviation equal to the $1\sigma$ error of the measured $logv_{rot}$.
    \item
    For each galaxy, we use the random $logv_{rot} $ obtained in the previous step to calculate the corresponding BTFR $logM_B$, using the best-fit slope and intercept of the real full dataset (first row of Table~\ref{table2}). We then obtain a random $logM_B$ for each galaxy from a Gaussian distribution with mean equal to the BTFR calculated $logM_B$ and standard deviation equal to the $1\sigma$ error of the measured $logM_B$.
    \item
    We repeat the above process 100 times, thereby generating 100 homogeneous Monte Carlo samples (HMCS) based on the SPARC dataset.
    \item
    For each HMCS, we find the $\sigma$ distances $d_\sigma(D_c)$ and for each $D_c$, we find the mean $\sigma$ distance and its standard deviation over the 100 HMCS. We thus construct the orange region in Figure~\ref{fig2_new} (right panel). A~typical form of $d_\sigma(D_c)$ is shown as the orange line of  Figure~\ref{fig2_new} (left panel) selected from the 100 HMCS. 
\end{itemize}

\begin{figure}[H]
\includegraphics[width = 0.75\textwidth]{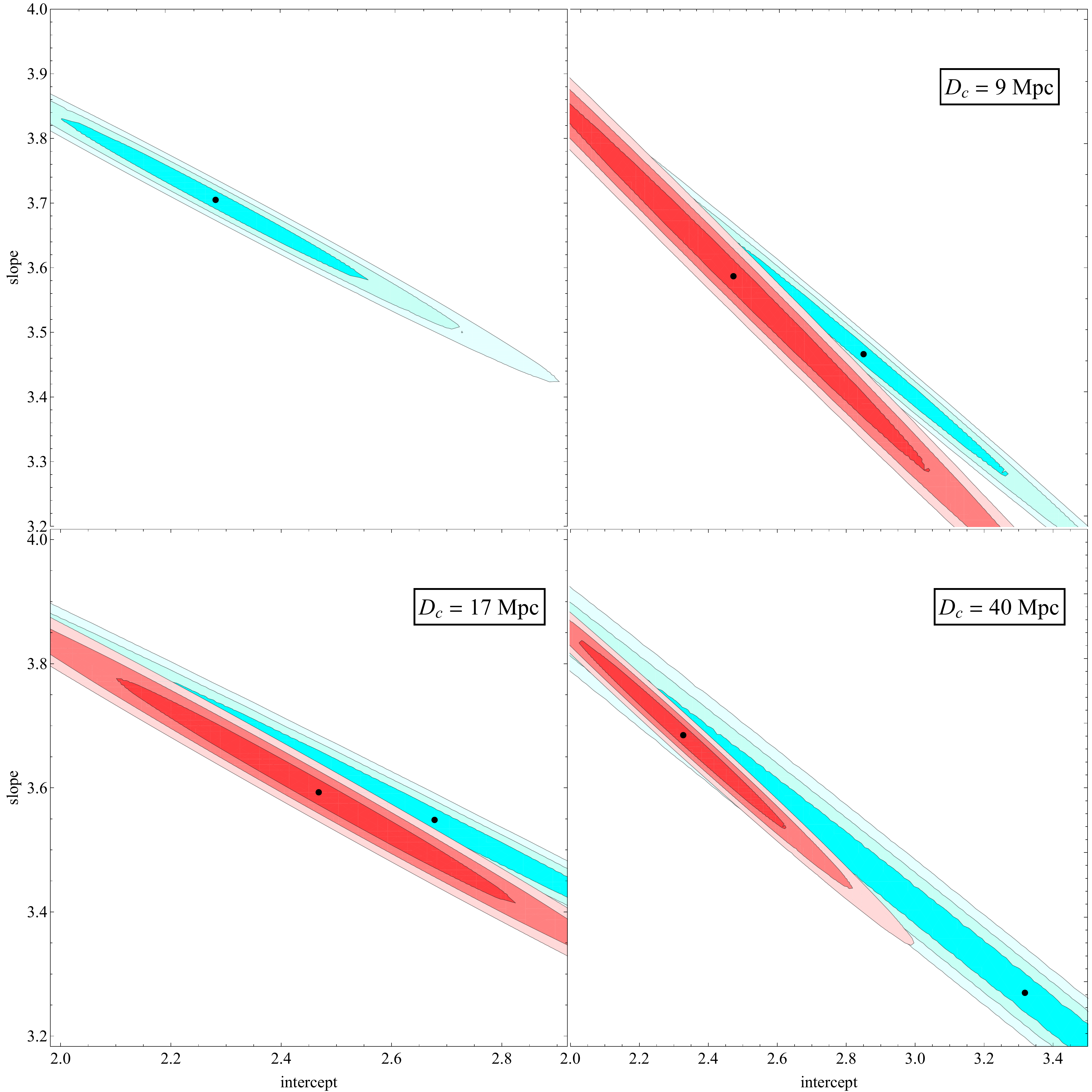}
\caption{The best-fit contours of the slope and intercept for the entire dataset, as~well for 3 different cases of split distance ($D_c$). The~red contours correspond to the dataset with galaxies that have a distance below $D_c$, whereas the cyan contours correspond to galaxies with distances above $D_c$.}
\label{fig2}
\end{figure}

Clearly, the forms of $d_\sigma(D_c)$ generated from the homogenized Monte Carlo samples  have the expected property to be confined mainly between $0\sigma$ and $2\sigma$ in contrast to the real measured sample, where $d_\sigma(D_c)$ extends up to $4\sigma$ or more. Thus, the real dataset is statistically distinct from a homogeneous BTFR~dataset.

The two maxima of $d_\sigma$ are more clearly illustrated in Figure~\ref{fig2}, where the likelihood contours are shown in the parameter space $s$ (slope)-$b$ (intercept) for the full sample (upper left panel) and for three pairs of subsamples $\Sigma_i$, including those corresponding to the peaks shown in Figure~\ref{fig1} ($D_c=17$ and $D_c=9$). For~both $d_\sigma$ maxima, the tension between the two best-fit points is mainly due to the different intercepts, while the values of the slope are very similar for the two subsamples. In~contrast, for~$D_c=40\rm\,Mpc$, where the $\sigma$ distance is much lower (about $1\sigma$, lower right panel), both the slope and the intercept differ significantly in magnitude but the statistical significance of this difference is low. Notice that the use of different statistics, such as the $1\sigma$ range of the best-fit intercept and slope shown in Table~\ref{table2}, or the level of likelihood contour overlap in Figure~\ref{fig2} would not reveal the tension between far and nearby subsamples. In~contrast, the $\sigma$-distance statistic demonstrates the effect and the Monte Carlo results of Figure~\ref{fig2_new} verify the fact that such a large $\sigma$-distance would be rare in the context of a homogeneous~sample.

The statistical significance of the different Tully--Fisher properties between near and far galaxies, which abruptly disappears for dividing distance $D_c\gtrsim 20$  Mpc, could be an unlikely statistical fluctuation, a~hint for systematics in the Tully--Fisher data\endnote{A possible source of systematics is the Malmquist bias, which would imply that the detected more distant galaxies are also more massive and may, therefore, display different slopes and intercepts in different mass bins~\cite{10.1093/mnras/stx458, 10.1093/mnrasl/slx134}.}, an~indication for an abrupt change in the galaxy evolution or a hint for a transition in the values of fundamental constants and, in particular, the strength of gravitational interactions $G_{\rm eff}$. The~best-fit values of the intercept and the slope for the cases shown in Figure~\ref{fig2} are displayed in Table~\ref{table2} along with their $1\sigma$ errors. 

\begin{figure}[H]
\includegraphics[width = 0.75 \textwidth]{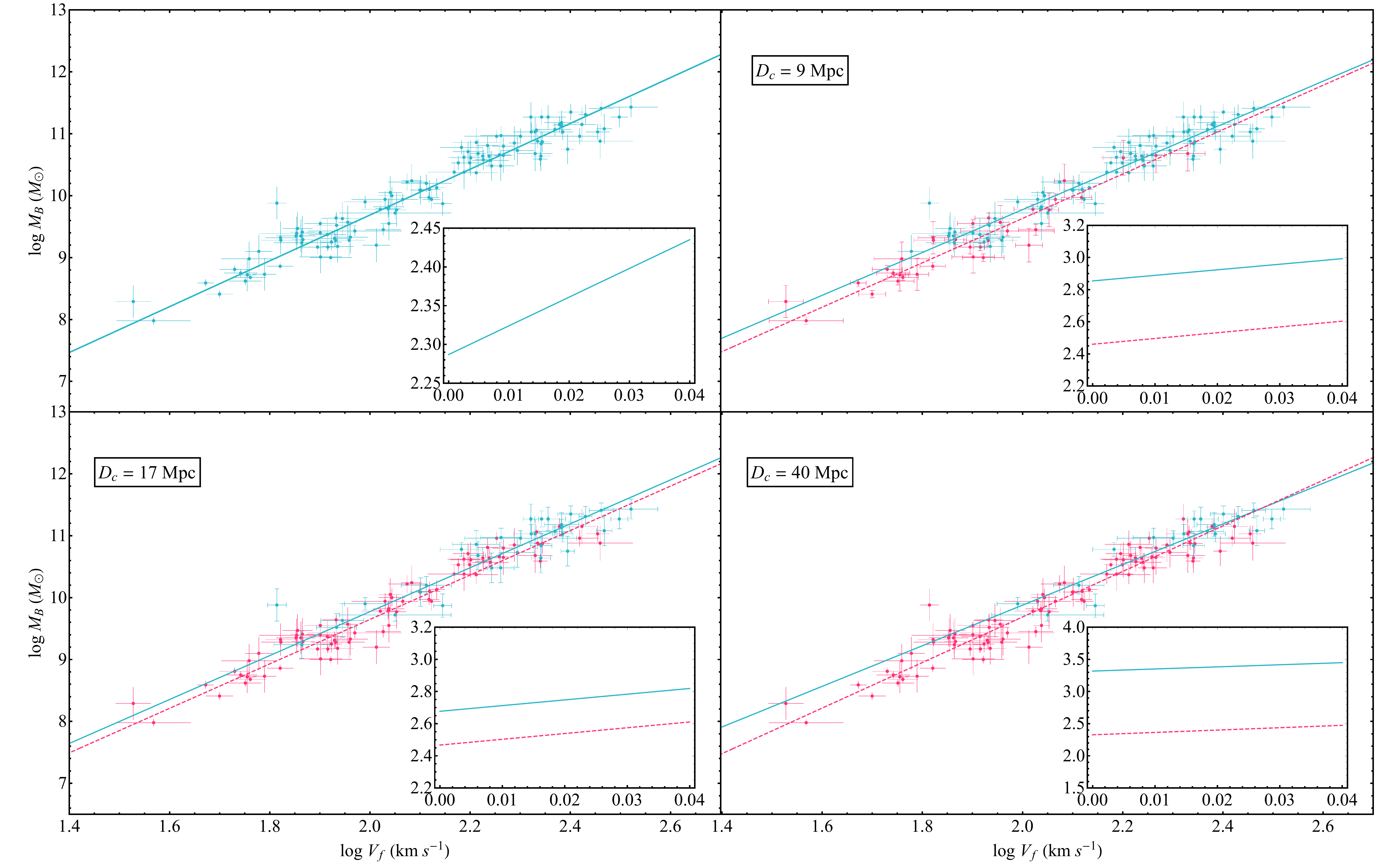}
\caption{The best-fit lines corresponding to the best fit slope and intercept parameters of the whole galaxy dataset as~well as each of the 2 datasets produced for 4 different split distances ($D_c$). The~red dashed line and datapoints correspond to the data below $D_c$, and~the cyan ones belong to the data over $D_c$ for each~case.}
\label{fig3}
\end{figure}

The best-fit $log M_B - log v_{rot}$ lines corresponding to Equation~\eqref{logbtfr} for the near--far galactic subsamples are shown in Figure~\ref{fig3}, superimposed with the datapoints (red/blue correspond to near/far galaxies). The~full dataset corresponds to the upper-left panel. The~difference between the two lines for $D_c=9\rm\,Mpc$ and $D_c=17\rm\,Mpc$ is evident, even though their slopes are very similar. The~statistical significance of this difference disappears for larger values of the splitting distance (\eg~$D_c=40\rm\,Mpc$), even though the slopes of the two lines become significantly different in this~case.

\begin{figure}[H]
\includegraphics[width = 0.43 \textwidth]{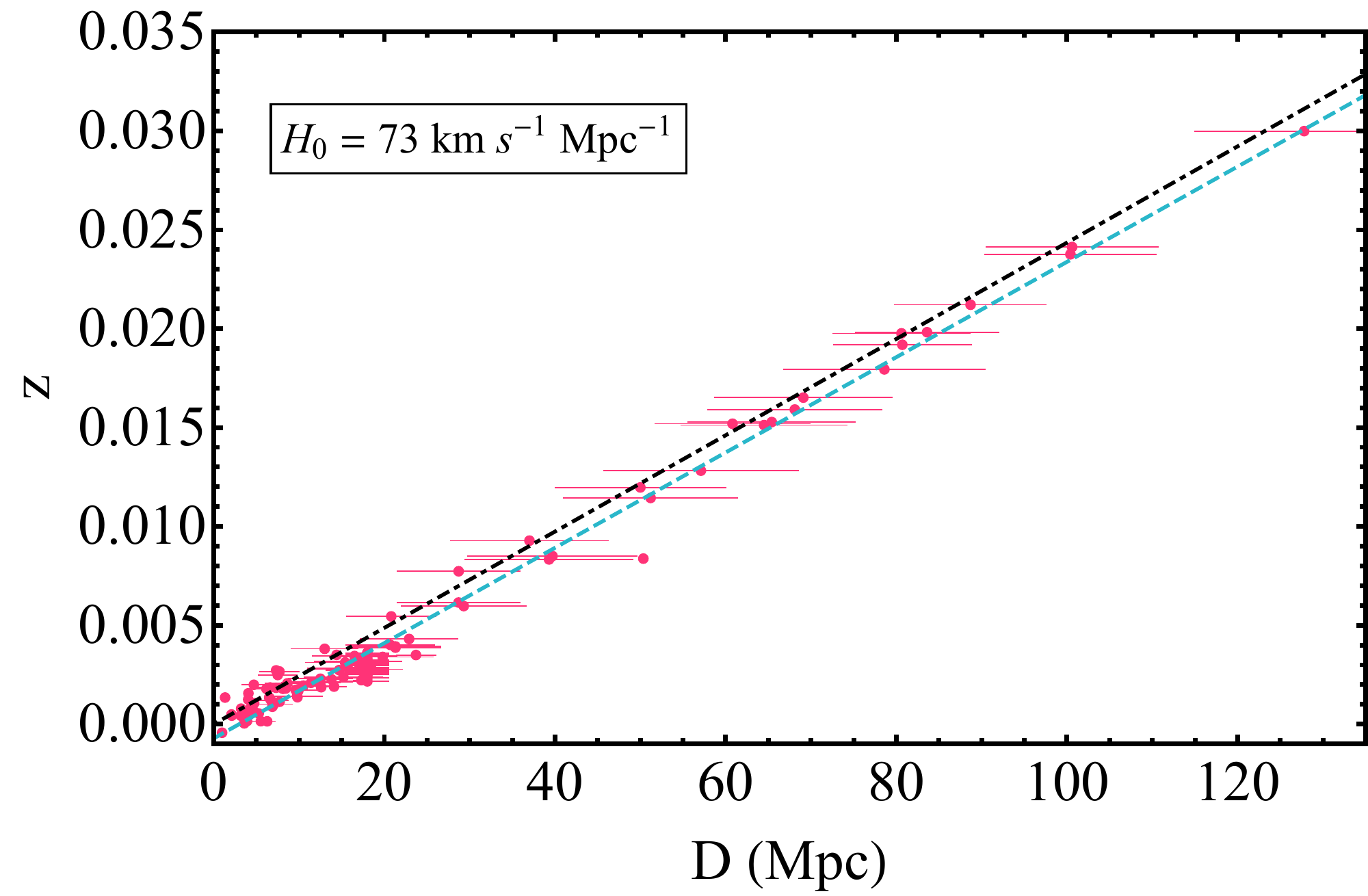}
\caption{The distances alongside their errorbars versus the redshifts of each galaxy in our compilation. The~blue dashed line corresponds to the best fit line, and~the black dot-dashed one is produced by Equation~\eqref{z_D} for $H_0 = 73\; \rm km\; s^{-1} \; Mpc^{-1}$.}
\label{fig4}
\end{figure}

\textls[-25]{The Hubble diagram of the considered dataset along with the best-fit line (black dot-dashed line) and the Hubble blue dashed line ($z \approx \frac{D}{c}H_0 \label{z_D}$) corresponding to\mbox{ $H_0 = 73\rm\,km\; s^{-1}\rm\, Mpc^{-1}$}} is shown in Figure~\ref{fig4}. The~distances to galaxies beyond 20 Mpc are determined using the Hubble flow with $H_0=73$ km/sec Mpc, and thus, there is no effect of their peculiar velocities. Galaxies closer than about $D\simeq 20\rm\,Mpc$ are clearly not in the Hubble flow and their redshift is affected significantly by their in-falling peculiar velocities, which tend to reduce their cosmological redshifts. The~detected transitions at about 9 Mpc and 17~Mpc correspond to cosmological redshifts of $z\lesssim 0.005$, which is lower than the transition redshift required for the resolution of the Hubble tension ($z_t\simeq 0.07$ is the upper redshift of SnIa--Cepheid host galaxies).

In the context of the above-described analysis, we have ignored the possible systematic uncertainties induced on the estimated baryonic masses $M_B$, due to systematic uncertainties in the measurement of galactic distances. In~particular, different sub-samples of galaxies in the SPARC database are affected by different systematic uncertainties.  The~SPARC sample includes galaxies with both direct and indirect distance measurements. Direct distance measurements are based on standard candles (Cepheids and Tip of Red Giant stars), while indirect measurements are based  on the Hubble flow with Virgocentric infall correction. Systematic uncertainties of indirectly measured distances affecting mainly galaxies beyond 15~Mpc are due to uncertainties in the Hubble constant $H_0$ and in the a Virgocentric infall model. $H_0=73\rm\, km/s/Mpc$ is assumed in estimating the distances of the Hubble flow subsample of the SPARC sample along with the Virgocentric infall model used to correct the Hubble flow distances. The~anticipated shift in $log M_B$ due to an incorrect assumption of the $H_0$ value and/or the Virgocentric infall model is anticipated to be of the order of $0.1$~dex, assuming a $5\%$ change in $H_0$ and a scaling of the estimated value of $M_B$ with distance $D$ as $M_B~D^{-2}$.

\begin{figure}[H]
\includegraphics[width =0.7\textwidth]{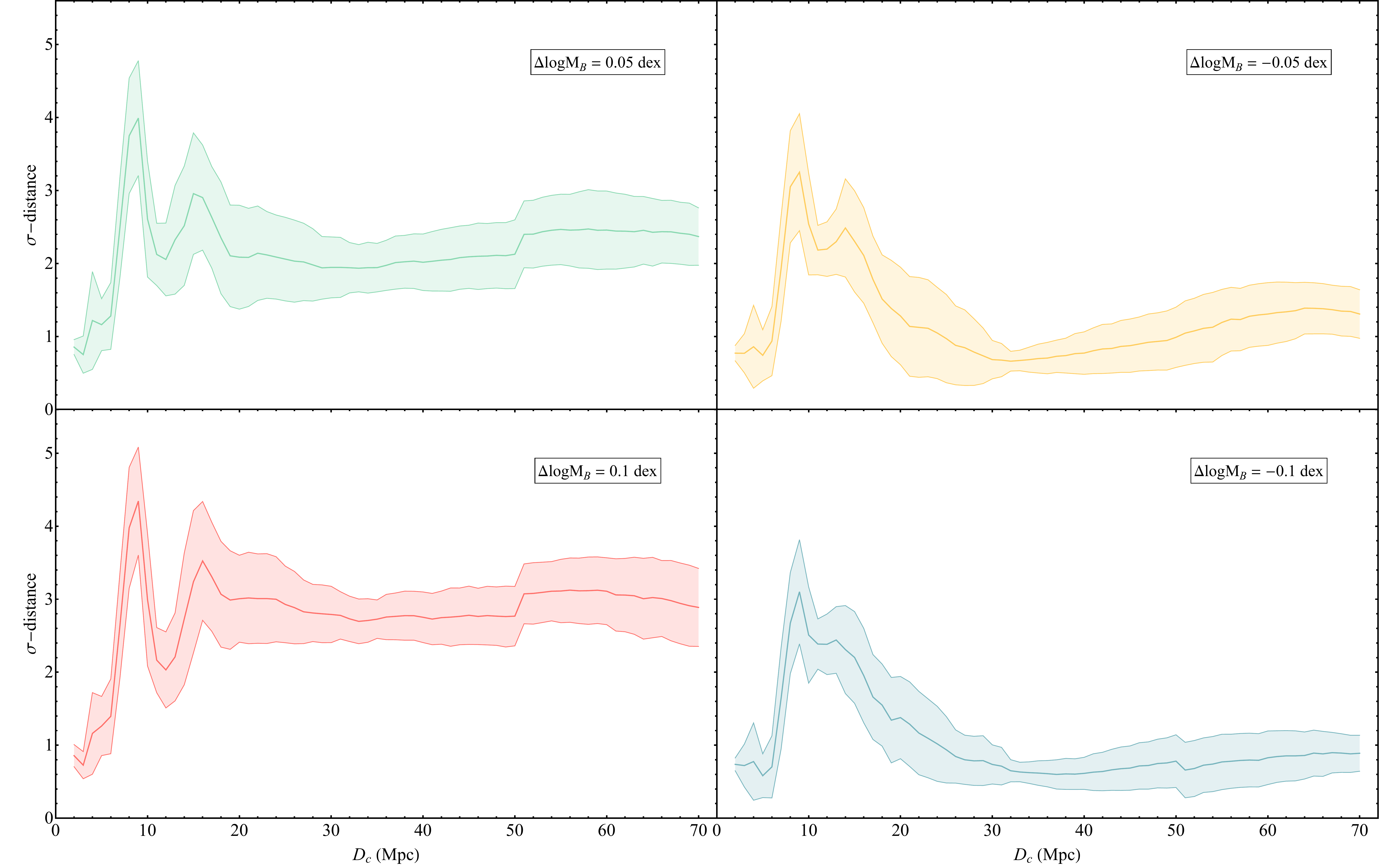}
\caption{The $68\%$ range of the $\sigma$-distances versus the split distances $D_c$ produced by a Monte Carlo simulation of 100 sample datasets. The~simulations are performed for different values of the shift $\Delta log M_B$, which represents the possible systematic errors present in the datapoints whose distances are calculated using the Hubble flow, assuming $H_0=73\rm\, km\, s^{-1}\, Mpc^{-1}$, and~correcting for Virgo-centric infall. The~same characteristic two-peak structure remains for all shifts considered, indicating the robust nature of the identified~effect.}
\label{figDM}
\end{figure} 

\begin{figure}[H]
\includegraphics[width = 0.73\textwidth]{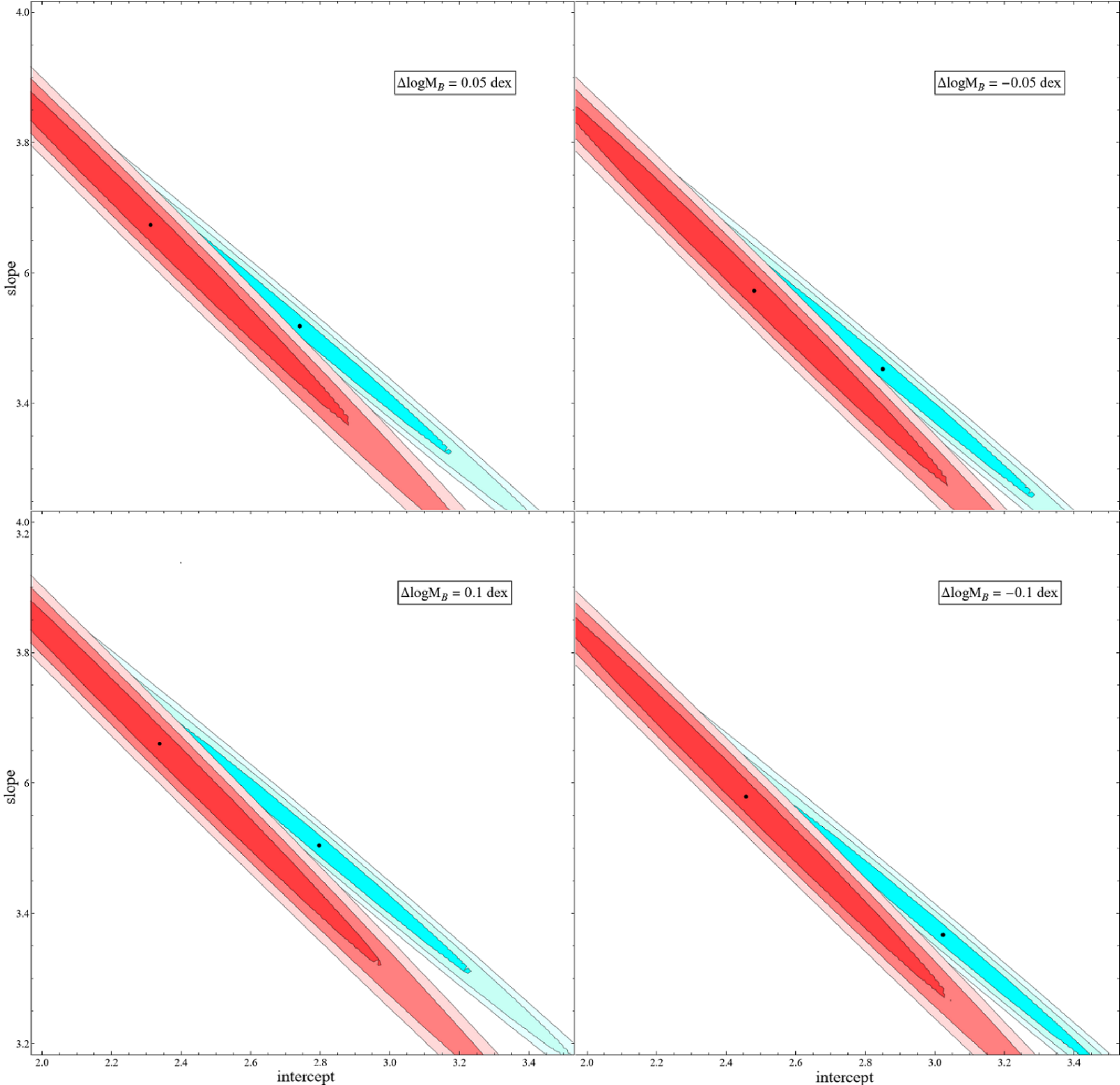}
\caption{The likelihood contours of the slope and intercept for a sample splitting distance $D_c = 9$~Mpc corresponding to the different values of the systematic shift $\Delta log M_B$ shown in Figure~\ref{figDM}. The~red contours correspond to the dataset with galaxies that have distance below 9~Mpc, whereas the cyan contours correspond to galaxies with distances above 9  Mpc. The~$\sigma$ distance between the two best fits varies between $3\sigma$ and $4\sigma$.}
\label{figcont9Mpc}
\end{figure}

Thus, the~identified mismatch of the Tully--Fisher parameters between low- and high-distance subsamples could, in principle, be due to such a systematic uncertainty of the galactic baryonic masses of Hubble flow galaxies. In~order to examine this possibility, we have constructed new Monte Carlo samples where we not only vary randomly the distances but also add a fixed shift of $\Delta log M_B$ along the vertical axis (mass) for all the datapoints where the mass is estimated using the Hubble flow with $H_0=73\rm\,km/s/Mpc$. The~distances of these points are calculated using the Hubble flow, assuming \mbox{$H_0=73\rm\,km\, s^{-1}\, Mpc^{-1}$}, and~correcting for Virgo-centric infall. We have considered four cases of systematic shifts (fixed values of $\Delta log M_B$):$-0.1$ dex, $-0.05$ dex, $+0.05$~dex and $+0.1$ dex.  The~results for the $\sigma$-distance ranges in terms of the splitting distance $D_c$ for each one of the above four cases are shown in Figure~\ref{figDM}. The~corresponding likelihood contours for the subsamples  corresponding to $D_c=9\rm\,Mpc$ (maximum mismatch) are shown in Figure~\ref{figcont9Mpc}. Clearly, the mismatch features at $D_c=9\rm\,Mpc$ and $D_c=17\rm\,Mpc$ remain in all four cases that explore this type of systematic uncertainty. In~particular, the~9~Mpc peak height varies from about $4\sigma$ for $\Delta log M_B=0.1$ dex to about $3\sigma$ for $\Delta log M_B=-0.1$ dex. We thus conclude that this type of systematic uncertainty is unable to wash out the mismatch effect we have~identified.


If the intercepts' transitions are interpreted as being due to a transition in $G_{\rm eff}$, we can use Equation~(\ref{abg}) along with the observed intercept transition amplitude shown in Table~\ref{table2} to identify the magnitude and sign of the corresponding $G_{\rm eff}$ transition. The~intercept transition at $D_c=17Mpc$ indicated in Table~\ref{table2} corresponds to the following: \be
\Delta \;log\; A_B\equiv log\; A_B^> - log\; A_B^<\simeq 0.2
\label{deltaloga}
\ee
Since $A_B$ is found to be higher at larger distances (early times), $G_{\rm eff}$ should be lower, due to Equation~(\ref{abg}).  The~corresponding fractional change in $G_{\rm eff}$ is easily obtained by differentiating the logarithmic form of Equation~(\ref{abg})  as follows:
\be 
\Delta \;log\; A_B=\frac{\Delta A_B}{A_B}=-2\frac{\Delta G_{\rm eff}}{G_{\rm eff}}\implies \frac{\Delta G_{\rm eff}}{G_{\rm eff}}\simeq -0.1 
\ee
This sign (weaker gravity at early times) and magnitude of the $G_{\rm eff}$ transition is consistent with the gravitational transition required for the resolution of the Hubble and growth tensions in the context of the mechanism of Ref.~\cite{Marra:2021fvf}.

\section{Conclusions - Discussion}\label{sec3}

We used a specific statistic on a robust dataset of 118 Tully--Fisher datapoints to demonstrate the existence of evidence for a transition in the evolution of BTFR. This evidence was verified by a wide range of Monte Carlo simulations that compare the real dataset with corresponding homogenized datasets constructed using the BTFR. It indicates a transition of the best-fit values of BTFR parameters, which is small in magnitude but~appears at a level of statistical significance of more than $3\sigma$. It corresponds to a transition of the intercept of the BTFR at a distance of $D_c\simeq 9\rm\,Mpc$ and/or at $D_c\simeq 17\rm\,Mpc$ (about 80 million years ago or less). Such a transition could be interpreted as a systematic effect or as a transition of the effective Newton constant with a $10\%$ lower value at early times, with the transition taking place about 80 million years ago or less. The~amplitude and sign of the gravitational transition are consistent with a recently proposed mechanism for the resolution of the Hubble and growth tensions~\cite{Marra:2021fvf,Alestas:2020zol}. However, the~time of the transition is about 60 million years later than the time suggested by the above mechanism  (100--150 million years ago corresponding to $D_c\simeq$ 30--40 Mpc and $z\simeq$ 0.007--0.01).

The effect shown in our analysis could be attributed to causes other than a gravitational transition. One such possible cause would be the presence of systematic errors affecting the estimate of galactic masses or rotation velocities for particular distance ranges. Even if this is the case, it is important to point out these inhomogeneities, which may require further analysis to identify their origin. Alternatively, if~the causes of the detected mismatch are physical, they could also be due to variation of conventional galaxy formation mechanisms, which may involve other types of modifications of gravitational physics (\eg~effects of MOND gravity). {The BTFR is an observationally tight empirical correlation and has therefore been used as a test of various modified gravity models (Refs.~\cite{Nojiri:2017ncd,Nojiri:2010wj,Clifton:2011jh} offer comprehensive reviews on the cosmological implications of such models), including modified Newtonian dynamics (MOND) \cite{1983ApJ...270..371M,McGaugh_2012} and Grumiller modified gravity~\cite{Ghosh:2021few}.These models have been shown to be consistent with BTFR for specific values of their acceleration parameters. The~BTFR has also been used as a test of the properties of Cold Dark Matter and galaxy formation mechanisms in the context of \lcdm~\cite{2008ApJ...682..861B,2010Natur.463..203G}.}

An interesting effect in the direction of the one observed in our analysis was also reported in Ref.~\cite{Mortsell:2021nzg}. There, the authors found a transition of the Cepheid magnitude behavior in the range of 10--20 Mpc, which could explain the Hubble tension (see Figure~4 of Ref.~\cite{Mortsell:2021nzg}). The~authors claimed that this transition is probably due to dust property variation, but there is currently a debate on the actual cause of this~mismatch.

An important extension of this analysis is the search for similar transition signals and constraints in other types of astrophysical and geophysical--climatological  data of Earth paleontology. For~example, a wide range of solar system anomalies were discussed in Ref.~\cite{Iorio:2014roa}, which could be revisited in the context of the gravitational transition hypothesis. Of~particular interest, for example, is the 'Faint young Sun paradox'~\cite{2012RvGeo..50G2006F}, which involves an inconsistency between geological findings and solar models about the temperature of the Earth about 4 billion years ago. Another interesting extension of this study would be the use of alternative methods for the identification of transition-like features in the data, \eg~the use of a Bayesian analysis tool, such as the internal robustness described in Refs.~\cite{2013MNRAS.430.1867A,Heneka:2013hka}.

Alternatively, other astrophysical relations that involve gravitational physics, such as the Faber--Jackson relation between intrinsic luminosity and velocity dispersion of elliptical galaxies or the Cepheid star period--luminosity relation, could also be screened for similar types of transitions as in the case of BTFR. For~example, the~question to address in the Cepheid case would be the following: `What constraints can be imposed on a transition-type evolution of the absolute magnitude ($M_v$)-period ($P$) relation of Population I Cepheid stars?' This relation may be written  as follows:
\be
M_v=s \; (logP-1)+b
\label{cephrel}
\ee
where $s=-2.43\pm 0.12$ and $b=-4.05\pm 0.02$ \cite{Benedict:2006cp,Benedict:2002ki}.

In conclusion, the low $z$ gravitational transition hypothesis is weakly constrained in the context of current studies but it could lead to the resolution of important cosmological tensions of the standard \lcdm model. We have demonstrated the existence of hints for such a transition in the evolution of the Tully--Fisher~relation.










\vspace{6pt}

\authorcontributions{LP contributed in the conceptualization, the methodology, the writing, as well as the general supervision of the project. GA contributed in the formal data analysis and the writing. IA contributed in the data curation and the literature investigation. All authors have read and agreed to the published version of the manuscript.} 


\funding{The research of LP and GA is co-financed by Greece and the European Union (European Social Fund---ESF) through the Operational Program ``Human Resources Development, Education and Lifelong Learning 2014-2020'' in the context of the project  MIS 5047648.}



\dataavailability{The numerical files for the reproduction of the figures can be found in the \href{https://github.com/GeorgeAlestas/Tully_Fisher_Transition}{Tully\_Fisher\_Transition} Github repository under the MIT license (Accessed date: 26/09/2021).}

\acknowledgments{We  thank Savvas Nesseris and Valerio Marra for their useful comments and suggestions. This research has made use of the SIMBAD database~\cite{2000A&AS..143....9W}, operated at CDS, Strasbourg,~France.}

\conflictsofinterest{The authors declare no conflict of~interest.}


\appendixtitles{yes} 

\appendixstart
\appendix
\section{Dataset of Galaxies~Used}\label{secA}
\textls[-5]{The following is the robust dataset of galaxies used in the analysis. We have used a compilation of 118 datapoints from Refs.~\cite{2001ApJ...563..694V,Walter_2008,2019MNRAS.484.3267L,2016ApJ...816L..14L}, for~which $M_B, v_{rot}$ and $D$ were~available.}

\begin{specialtable}[H]

\caption{The robust compilation of galaxy data found in Refs.~\cite{2001ApJ...563..694V,Walter_2008,2019MNRAS.484.3267L,2016ApJ...816L..14L}.}\label{table3}

\setlength{\cellWidtha}{\columnwidth/7-2\tabcolsep+0.2in}
\setlength{\cellWidthb}{\columnwidth/7-2\tabcolsep+0in}
\setlength{\cellWidthc}{\columnwidth/7-2\tabcolsep-0.05in}
\setlength{\cellWidthd}{\columnwidth/7-2\tabcolsep-0.05in}
\setlength{\cellWidthe}{\columnwidth/7-2\tabcolsep-0.0in}
\setlength{\cellWidthf}{\columnwidth/7-2\tabcolsep-0.1in}
\setlength{\cellWidthg}{\columnwidth/7-2\tabcolsep-0in}

\scalebox{1}[1]{\begin{tabularx}{\columnwidth}{>{\PreserveBackslash\centering}m{\cellWidtha}>{\PreserveBackslash\centering}m{\cellWidthb}>{\PreserveBackslash\centering}m{\cellWidthc}>{\PreserveBackslash\centering}m{\cellWidthd}>{\PreserveBackslash\centering}m{\cellWidthe}>{\PreserveBackslash\centering}m{\cellWidthf}>{\PreserveBackslash\centering}m{\cellWidthg}}
\toprule

\textbf{Galaxy Name} & \boldmath$Log v_{rot}$ & \boldmath$\sigma_{Log v_{rot}}$ & \boldmath$Log M_B$   & \boldmath$\sigma_{Log M_B}$  &\boldmath $D$  & \boldmath$\sigma_D$  \\\midrule 

& \textbf{(km/s)} &  \textbf{(km/s)} &  \boldmath\textbf{($M_\odot$)} & \boldmath\textbf{($M_\odot$) }&\boldmath\textbf{ ($Mpc$) }&\boldmath \textbf{ ($Mpc$)} \\
\midrule 

\text{D631-7} & 1.76 & 0.03 & 8.68 & 0.05 & 7.72 & 0.39 \\
\text{DDO154} & 1.67 & 0.02 & 8.59 & 0.06 & 4.04 & 0.2 \\
\text{DDO161} & 1.82 & 0.03 & 9.32 & 0.26 & 7.5 & 2.25 \\
 \text{DDO168} & 1.73 & 0.03 & 8.81 & 0.06 & 4.25 & 0.21 \\
 \text{DDO170} & 1.78 & 0.03 & 9.1 & 0.26 & 15.4 & 4.62 \\
 \text{ESO079-G014} & 2.24 & 0.01 & 10.48 & 0.24 & 28.7 & 7.17 \\
 \text{ESO116-G012} & 2.04 & 0.02 & 9.55 & 0.27 & 13. & 3.9 \\
 \text{ESO563-G021} & 2.5 & 0.02 & 11.27 & 0.16 & 60.8 & 9.1 \\
 \text{F568-V1} & 2.05 & 0.11 & 9.72 & 0.1 & 80.6 & 8.06 \\
 \text{F571-8} & 2.15 & 0.02 & 9.87 & 0.19 & 53.3 & 10.7 \\
 \text{F574-1} & 1.99 & 0.04 & 9.9 & 0.1 & 96.8 & 9.68 \\
 \text{F583-1} & 1.93 & 0.04 & 9.52 & 0.22 & 35.4 & 8.85 \\
 \text{IC2574} & 1.82 & 0.04 & 9.28 & 0.06 & 3.91 & 0.2 \\
 \text{IC4202} & 2.38 & 0.02 & 11.03 & 0.13 & 100.4 & 10. \\
 \text{KK98-251} & 1.53 & 0.03 & 8.29 & 0.26 & 6.8 & 2.04 \\
 \text{NGC0024} & 2.03 & 0.04 & 9.45 & 0.09 & 7.3 & 0.36 \\
 \text{NGC0055} & 1.93 & 0.03 & 9.64 & 0.08 & 2.11 & 0.11 \\
 \text{NGC0100} & 1.94 & 0.04 & 9.63 & 0.27 & 18.45 & 0.2 \\
 \text{NGC0247} & 2.02 & 0.04 & 9.78 & 0.08 & 3.7 & 0.19 \\
 \text{NGC0289} & 2.21 & 0.05 & 10.86 & 0.22 & 20.8 & 5.2 \\
 \text{NGC0300} & 1.97 & 0.09 & 9.43 & 0.08 & 2.08 & 0.1 \\
  \text{NGC0801} & 2.34 & 0.01 & 11.27 & 0.13 & 80.7 & 8.07 \\
 \text{NGC0891} & 2.33 & 0.01 & 10.88 & 0.11 & 9.91 & 0.5 \\
 \text{NGC1003} & 2.04 & 0.02 & 10.05 & 0.26 & 11.4 & 3.42 \\
 \text{NGC1090} & 2.22 & 0.02 & 10.68 & 0.23 & 37. & 9.25 \\
 \text{NGC2403} & 2.12 & 0.02 & 9.97 & 0.08 & 3.16 & 0.16 \\
 \text{NGC2683} & 2.19 & 0.03 & 10.62 & 0.11 & 9.81 & 0.49 \\
 \text{NGC2841} & 2.45 & 0.02 & 11.03 & 0.13 & 14.1 & 1.4 \\
 \text{NGC2903} & 2.27 & 0.02 & 10.65 & 0.28 & 6.6 & 1.98 \\

 \bottomrule
\end{tabularx}}
\end{specialtable}

\begin{specialtable}[H]\ContinuedFloat
\caption{{\em Cont.}}
\setlength{\cellWidtha}{\columnwidth/7-2\tabcolsep+0.2in}
\setlength{\cellWidthb}{\columnwidth/7-2\tabcolsep+0in}
\setlength{\cellWidthc}{\columnwidth/7-2\tabcolsep-0.05in}
\setlength{\cellWidthd}{\columnwidth/7-2\tabcolsep-0.05in}
\setlength{\cellWidthe}{\columnwidth/7-2\tabcolsep-0.0in}
\setlength{\cellWidthf}{\columnwidth/7-2\tabcolsep-0.1in}
\setlength{\cellWidthg}{\columnwidth/7-2\tabcolsep-0in}

\scalebox{1}[1]{\begin{tabularx}{\columnwidth}{>{\PreserveBackslash\centering}m{\cellWidtha}>{\PreserveBackslash\centering}m{\cellWidthb}>{\PreserveBackslash\centering}m{\cellWidthc}>{\PreserveBackslash\centering}m{\cellWidthd}>{\PreserveBackslash\centering}m{\cellWidthe}>{\PreserveBackslash\centering}m{\cellWidthf}>{\PreserveBackslash\centering}m{\cellWidthg}}
\toprule

\textbf{Galaxy Name} & \boldmath$Log v_{rot}$ & \boldmath$\sigma_{Log v_{rot}}$ & \boldmath$Log M_B$   & \boldmath$\sigma_{Log M_B}$  &\boldmath $D$  & \boldmath$\sigma_D$  \\\midrule 

& \textbf{(km/s)} &  \textbf{(km/s)} &  \boldmath\textbf{($M_\odot$)} & \boldmath\textbf{($M_\odot$) }&\boldmath\textbf{ ($Mpc$) }&\boldmath \textbf{ ($Mpc$)} \\
\midrule

 \text{NGC2915} & 1.92 & 0.04 & 9. & 0.06 & 4.06 & 0.2 \\
 \text{NGC2976} & 1.93 & 0.05 & 9.28 & 0.11 & 3.58 & 0.18 \\
 \text{NGC2998} & 2.32 & 0.02 & 11.03 & 0.15 & 68.1 & 10.2 \\
 \text{NGC3109} & 1.82 & 0.03 & 8.86 & 0.06 & 1.33 & 0.07 \\
 \text{NGC3198} & 2.18 & 0.01 & 10.53 & 0.11 & 13.8 & 1.4 \\
 \text{NGC3521} & 2.33 & 0.03 & 10.68 & 0.28 & 7.7 & 2.3 \\
 \text{NGC3726} & 2.23 & 0.03 & 10.64 & 0.15 & 18. & 2.5 \\
 \text{NGC3741} & 1.7 & 0.03 & 8.41 & 0.06 & 3.21 & 0.17 \\
 \text{NGC3769} & 2.07 & 0.04 & 10.22 & 0.14 & 18. & 2.5 \\
 \text{NGC3877} & 2.23 & 0.02 & 10.58 & 0.16 & 18. & 2.5 \\
 \text{NGC3893} & 2.25 & 0.04 & 10.57 & 0.15 & 18. & 2.5 \\
 \text{NGC3917} & 2.13 & 0.02 & 10.13 & 0.15 & 18. & 2.5 \\
 \text{NGC3949} & 2.21 & 0.04 & 10.37 & 0.15 & 18. & 2.5 \\
 \text{NGC3953} & 2.34 & 0.02 & 10.87 & 0.16 & 18. & 2.5 \\
 \text{NGC3972} & 2.12 & 0.02 & 9.94 & 0.15 & 18. & 2.5 \\
 \text{NGC3992} & 2.38 & 0.02 & 11.13 & 0.13 & 23.7 & 2.3 \\
 \text{NGC4010} & 2.1 & 0.02 & 10.09 & 0.14 & 18. & 2.5 \\
 \text{NGC4013} & 2.24 & 0.02 & 10.64 & 0.16 & 18. & 2.5 \\
 \text{NGC4051} & 2.2 & 0.03 & 10.71 & 0.16 & 18. & 2.5 \\
 \text{NGC4085} & 2.12 & 0.02 & 10.1 & 0.15 & 18. & 2.5 \\
 \text{NGC4088} & 2.24 & 0.02 & 10.81 & 0.15 & 18. & 2.5 \\
 \text{NGC4100} & 2.2 & 0.02 & 10.53 & 0.15 & 18. & 2.5 \\
 \text{NGC4138} & 2.17 & 0.05 & 10.38 & 0.16 & 18. & 2.5 \\
 \text{NGC4157} & 2.27 & 0.02 & 10.8 & 0.15 & 18. & 2.5 \\
 \text{NGC4183} & 2.04 & 0.03 & 10. & 0.14 & 18. & 2.5 \\
 \text{NGC4217} & 2.26 & 0.02 & 10.66 & 0.16 & 18. & 2.5 \\
 \text{NGC4559} & 2.08 & 0.02 & 10.24 & 0.27 & 7.31 & 0.2 \\
 \text{NGC5005} & 2.42 & 0.04 & 10.96 & 0.13 & 16.9 & 1.5 \\
 \text{NGC5033} & 2.29 & 0.01 & 10.85 & 0.27 & 15.7 & 4.7 \\
 \text{NGC5055} & 2.26 & 0.03 & 10.96 & 0.1 & 9.9 & 0.5 \\
 \text{NGC5371} & 2.32 & 0.02 & 11.27 & 0.24 & 39.7 & 9.92 \\
 \text{NGC5585} & 1.96 & 0.02 & 9.57 & 0.27 & 7.06 & 2.12 \\
 \text{NGC5907} & 2.33 & 0.01 & 11.06 & 0.1 & 17.3 & 0.9 \\
 \text{NGC5985} & 2.47 & 0.02 & 11.08 & 0.24 & 50.35 & 0.2 \\
 \text{NGC6015} & 2.19 & 0.02 & 10.38 & 0.27 & 17. & 5.1 \\
 \text{NGC6195} & 2.40 & 0.03 & 11.35 & 0.13 & 127.8 & 12.8 \\
 \text{NGC6503} & 2.07 & 0.01 & 9.94 & 0.09 & 6.26 & 0.31 \\
 \text{NGC6674} & 2.38 & 0.03 & 11.18 & 0.19 & 51.2 & 10.2 \\
 \text{NGC6946} & 2.20 & 0.04 & 10.61 & 0.28 & 5.52 & 1.66 \\
 \text{NGC7331} & 2.38 & 0.01 & 11.15 & 0.13 & 14.7 & 1.5 \\
 \text{NGC7814} & 2.34 & 0.01 & 10.59 & 0.11 & 14.4 & 0.72 \\
 \text{UGC00128} & 2.12 & 0.05 & 10.2 & 0.14 & 64.5 & 9.7 \\
 \text{UGC00731} & 1.87 & 0.02 & 9.41 & 0.26 & 12.5 & 3.75 \\
 \text{UGC01281} & 1.75 & 0.03 & 8.75 & 0.06 & 5.27 & 0.1 \\
 \text{UGC02259} & 1.94 & 0.03 & 9.18 & 0.26 & 10.5 & 3.1 \\
 \text{UGC02487} & 2.52 & 0.05 & 11.43 & 0.16 & 69.1 & 10.4 \\
 \text{UGC02885} & 2.46 & 0.02 & 11.41 & 0.12 & 80.6 & 8.06 \\
 \text{UGC02916} & 2.26 & 0.04 & 10.97 & 0.15 & 65.4 & 9.8 \\
 \text{UGC02953} & 2.42 & 0.03 & 11.15 & 0.28 & 16.5 & 4.95 \\
 \text{UGC03205} & 2.34 & 0.02 & 10.84 & 0.2 & 50. & 10. \\
 \text{UGC03546} & 2.29 & 0.03 & 10.73 & 0.24 & 28.7 & 7.2 \\
 \text{UGC03580} & 2.10 & 0.02 & 10.09 & 0.23 & 20.7 & 5.2 \\
 \text{UGC04278} & 1.96 & 0.03 & 9.33 & 0.26 & 12.59 & 0.2 \\
 \text{UGC04325} & 1.96 & 0.03 & 9.28 & 0.27 & 9.6 & 2.88 \\
 \text{UGC04499} & 1.86 & 0.03 & 9.35 & 0.26 & 12.5 & 3.75 \\
 \text{UGC05253} & 2.33 & 0.04 & 11.03 & 0.23 & 22.9 & 5.72 \\

 \bottomrule
\end{tabularx}}
\end{specialtable}

\begin{specialtable}[H]\ContinuedFloat

\caption{{\em Cont.}}
\setlength{\cellWidtha}{\columnwidth/7-2\tabcolsep+0.2in}
\setlength{\cellWidthb}{\columnwidth/7-2\tabcolsep+0in}
\setlength{\cellWidthc}{\columnwidth/7-2\tabcolsep-0.05in}
\setlength{\cellWidthd}{\columnwidth/7-2\tabcolsep-0.05in}
\setlength{\cellWidthe}{\columnwidth/7-2\tabcolsep-0.0in}
\setlength{\cellWidthf}{\columnwidth/7-2\tabcolsep-0.1in}
\setlength{\cellWidthg}{\columnwidth/7-2\tabcolsep-0in}

\scalebox{1}[1]{\begin{tabularx}{\columnwidth}{>{\PreserveBackslash\centering}m{\cellWidtha}>{\PreserveBackslash\centering}m{\cellWidthb}>{\PreserveBackslash\centering}m{\cellWidthc}>{\PreserveBackslash\centering}m{\cellWidthd}>{\PreserveBackslash\centering}m{\cellWidthe}>{\PreserveBackslash\centering}m{\cellWidthf}>{\PreserveBackslash\centering}m{\cellWidthg}}
\toprule

\textbf{Galaxy Name} & \boldmath$Log v_{rot}$ & \boldmath$\sigma_{Log v_{rot}}$ & \boldmath$Log M_B$   & \boldmath$\sigma_{Log M_B}$  &\boldmath $D$  & \boldmath$\sigma_D$  \\\midrule 

& \textbf{(km/s)} &  \textbf{(km/s)} &  \boldmath\textbf{($M_\odot$)} & \boldmath\textbf{($M_\odot$) }&\boldmath\textbf{ (Mpc) }&\boldmath \textbf{ (Mpc)} \\
\midrule 
 \text{UGC05716} & 1.87 & 0.06 & 9.24 & 0.22 & 21.3 & 5.3 \\
 \text{UGC05721} & 1.9 & 0.04 & 9.01 & 0.26 & 6.18 & 1.85 \\
 \text{UGC05986} & 2.05 & 0.02 & 9.77 & 0.27 & 8.63 & 2.59 \\
 \text{UGC06399} & 1.93 & 0.03 & 9.31 & 0.14 & 18. & 2.5 \\
 \text{UGC06446} & 1.92 & 0.04 & 9.37 & 0.26 & 12. & 3.6 \\
 \text{UGC06614} & 2.3 & 0.11 & 10.96 & 0.12 & 88.7 & 8.87 \\
 \text{UGC06667} & 1.92 & 0.02 & 9.25 & 0.13 & 18. & 2.5 \\
 \text{UGC06786} & 2.34 & 0.02 & 10.64 & 0.24 & 29.3 & 7.32 \\
 \text{UGC06787} & 2.4 & 0.01 & 10.75 & 0.24 & 21.3 & 5.32 \\
 \text{UGC06818} & 1.85 & 0.04 & 9.35 & 0.13 & 18. & 2.5 \\
 \text{UGC06917} & 2.04 & 0.03 & 9.79 & 0.14 & 18. & 2.5 \\
 \text{UGC06923} & 1.90 & 0.03 & 9.4 & 0.14 & 18. & 2.5 \\
 \text{UGC06930} & 2.03 & 0.07 & 9.94 & 0.13 & 18. & 2.5 \\
 \text{UGC06983} & 2.04 & 0.03 & 9.82 & 0.13 & 18. & 2.5 \\
 \text{UGC07125} & 1.81 & 0.03 & 9.88 & 0.26 & 19.8 & 5.9 \\
 \text{UGC07151} & 1.87 & 0.02 & 9.29 & 0.08 & 6.87 & 0.34 \\
 \text{UGC07399} & 2.01 & 0.03 & 9.2 & 0.27 & 8.43 & 2.53 \\
 \text{UGC07524} & 1.9 & 0.03 & 9.55 & 0.06 & 4.74 & 0.24 \\
 \text{UGC07603} & 1.79 & 0.02 & 8.73 & 0.26 & 4.7 & 1.41 \\
 \text{UGC07690} & 1.76 & 0.06 & 8.98 & 0.27 & 8.11 & 2.43 \\
 \text{UGC08286} & 1.92 & 0.01 & 9.17 & 0.06 & 6.5 & 0.33 \\
 \text{UGC08490} & 1.9 & 0.03 & 9.17 & 0.11 & 4.65 & 0.53 \\
 \text{UGC08550} & 1.76 & 0.02 & 8.72 & 0.26 & 6.7 & 2. \\
 \text{UGC08699} & 2.26 & 0.03 & 10.48 & 0.24 & 39.3 & 9.82 \\
 \text{UGC09037} & 2.18 & 0.04 & 10.78 & 0.11 & 83.6 & 8.4 \\
 \text{UGC09133} & 2.36 & 0.04 & 11.27 & 0.19 & 57.1 & 11.4 \\
 \text{UGC10310} & 1.85 & 0.08 & 9.39 & 0.27 & 15.2 & 4.6 \\
 \text{UGC11455} & 2.43 & 0.01 & 11.31 & 0.16 & 78.6 & 11.8 \\
 \text{UGC11914} & 2.46 & 0.07 & 10.88 & 0.28 & 16.9 & 5.1 \\
 \text{UGC12506} & 2.37 & 0.03 & 11.07 & 0.11 & 100.6 & 10.1 \\
 \text{UGC12632} & 1.86 & 0.03 & 9.47 & 0.26 & 9.77 & 2.93 \\
 \text{UGCA442} & 1.75 & 0.03 & 8.62 & 0.06 & 4.35 & 0.22 \\
 \text{UGCA444} & 1.57 & 0.07 & 7.98 & 0.06 & 0.98 & 0.05 \\ \bottomrule
\end{tabularx}}
    \end{specialtable}

\begin{adjustwidth}{-4.6cm}{0cm}
\printendnotes[custom]
\end{adjustwidth}

\end{paracol}
\reftitle{References}


\externalbibliography{yes}



\end{document}